\documentclass[runningheads]{llncs}
\usepackage[T1]{fontenc}
\usepackage{graphicx,verbatim}
\usepackage{float}
\usepackage{amsmath}
\usepackage{amssymb}

\usepackage{graphicx}
\usepackage[colorlinks,urlcolor=blue]{hyperref}
\usepackage{color}
\usepackage{url}  

\usepackage{mathtools}
\usepackage{amsmath,amsfonts,amssymb}

\usepackage{tabularray}
\usepackage[misc]{ifsym}
\usepackage{booktabs}
\usepackage{upgreek}
\usepackage{enumitem}
\usepackage{bm}
\usepackage{censor}
\usepackage{textcomp}
\usepackage[english]{babel}
\usepackage{lastpage}
\usepackage[section]{placeins}
\usepackage{fancyvrb} 
\usepackage[dvipsnames,table]{xcolor}
\usepackage{fontawesome5}  

\usepackage{multirow}
\usepackage{makecell}
\usepackage{subfigure}
\usepackage{tabularx,ragged2e}
\usepackage{booktabs}
\usepackage{pifont}

\definecolor{LightCyan}{rgb}{0.88,1,1}
\definecolor{lightskyblue}{RGB}{225, 235, 240}

\definecolor{Gray}{gray}{0.90}
\definecolor{white}{rgb}{1.0, 1.0, 1.0}

\definecolor{Lightgreen}{RGB}{218, 246, 230 }

\begin{document}
%
\title{UNICON: UNIfied CONtinual Learning for Medical Foundational Models}
%

\titlerunning{UNICON: UNIfied CONtinual Learning}

\author{Mohammad Areeb Qazi, Munachiso S Nwadike, Ibrahim Almakky, Mohammad Yaqub and Numan Saeed}
\authorrunning{Qazi et al.}
\institute{Mohamed bin Zayed University of Artificial Intelligence, Abu Dhabi, UAE
\email{\{firstname.lastname\}@mbzuai.ac.ae}}

\maketitle              
\begin{abstract}

Foundational models are trained on extensive datasets to capture the general trends of a domain. However, in medical imaging, the scarcity of data makes pre-training for every domain, modality, or task challenging. Continual learning offers a solution by fine-tuning a model sequentially on different domains or tasks, enabling it to integrate new knowledge without requiring large datasets for each training phase. In this paper, we propose UNIfied CONtinual Learning for Medical Foundational Models (\textbf{UNICON}), a framework that enables the seamless adaptation of foundation models to diverse domains, tasks, and modalities. Unlike conventional adaptation methods that treat these changes in isolation, UNICON provides a unified, perpetually expandable framework. Through careful integration, we show that foundation models can dynamically expand across imaging modalities, anatomical regions, and clinical objectives without catastrophic forgetting or task interference. Empirically, we validate our approach by adapting a chest CT  foundation model initially trained for classification to a prognosis and segmentation task. Our results show improved performance across both additional tasks. Furthermore, we continually incorporated PET scans and achieved a 5\% improvement in Dice score compared to respective baselines. These findings establish that foundation models are not inherently constrained to their initial training scope but can evolve, paving the way toward generalist AI models for medical imaging.

\keywords{Continual Learning  \and Efficient Medical Imaging \and Multi-modal Adaptation}

\end{abstract}
\section{Introduction}

Medical imaging plays a crucial role in modern healthcare, providing essential visual insights that are key to diagnosing, monitoring, and treating a wide array of conditions \cite{litjens2017survey,giger2018machine,ranschaert2019artificial}. Despite its importance, collecting and curating medical imaging data is highly resource-intensive, with large hospitals generating approximately 100 terabytes of imaging data per year, which require significant storage, processing power, and infrastructure to effectively manage. \cite{duke2015healthdata,langer2011multisite,openmed2023bigdata}. As a result, it becomes difficult to share medical data. 

Now, with the introduction of Medical Foundational Models (FM) pre-trained on large-scale data, deep learning systems have demonstrated exceptional performance within their trained domains \cite{zhao2024biomedparse,wu2023towards,hamamci2024developing,liang20243d,zhang2024foundation}. These models are typically tailored to their specific domains by leveraging specialized datasets, architectures, and training objectives that align with medical imaging tasks. For instance, they are trained on radiology reports, CT scans, MRIs, or pathology slides, enabling them to capture domain-specific patterns and features that general-purpose models may overlook \cite{wu2023radfm,yu2024chief,taher2023eden}. However, the diversity of medical imaging modalities presents a significant challenge for FM. For instance, CT provides detailed cross-sectional images of anatomical structures, while PET visualizes metabolic processes, highlighting functional aspects of tissues. Each modality captures different aspects of human anatomy and physiology, leading to distinct data distributions. This variability can hinder the generalizability of FM across all imaging types, as they may struggle to perform consistently well on modalities not represented in their training data \cite{shi2024trustworthy}. 

Training FM for each medical imaging modality is not feasible due to the vast data requirements, which are often unavailable in healthcare. As a result, adaptation techniques have been employed to extend models across domains, classes, and tasks \cite{saadi2024pemma,qazi2024dynammo,zhang2023adapter,chen2024low,sun2024continually}. Studies have demonstrated the adaptation of classification models to incorporate new classes while maintaining their original task \cite{qazi2024continual,zhang2023adapter}, segmentation models to generalize across different anatomical regions \cite{chen2024low}, and report generation models to generate structured outputs for diverse medical findings \cite{sun2024continually}. Additionally, research has shown that models can be adapted to integrate multiple modalities over the same anatomical region \cite{sobirov2022automatic,saadi2024pemma}. 



Despite these advancements, existing approaches primarily focus on continual adaptation within a single task or domain, lacking a unified framework to extend foundation models across new tasks, imaging modalities, and anatomical regions. Moreover, unlike pretraining on curated datasets, real-world medical imaging data often varies in resolution and quality, posing additional challenges in adapting FM to diverse tasks and domains. This highlights the need for a comprehensive solution that enables seamless adaptation across all dimensions. To this end, we propose UNIfied CONtinual Learning for Medical Foundational Models (\textbf{UNICON}), a framework for the continual adaptation of a FM initially trained for a specific modality, task, and anatomical region, enabling it to generalize efficiently to new modalities, tasks, and anatomical regions. Unlike traditional continual learning and conventional domain adaptation techniques, our approach facilitates multi-step adaptation across diverse scenarios, ensuring scalability and robustness in medical imaging applications. Our contributions can be summarized as follows:

\begin{itemize}
    \item We introduce \textbf{UNICON}, a framework that enables foundation models to be continually extended across new tasks, modalities, and anatomical regions without performance degradation.
    \item We propose a method for adapting foundation models to arbitrary resolution sizes, allowing seamless generalization beyond their original training constraints.
    \item We demonstrate that medical foundation models can integrate new knowledge sequentially without catastrophic forgetting, reducing the need for separate specialized models.
\end{itemize}

\section{Methodology}

\begin{figure}[t]
    \centering
    \includegraphics[width=0.8\linewidth]{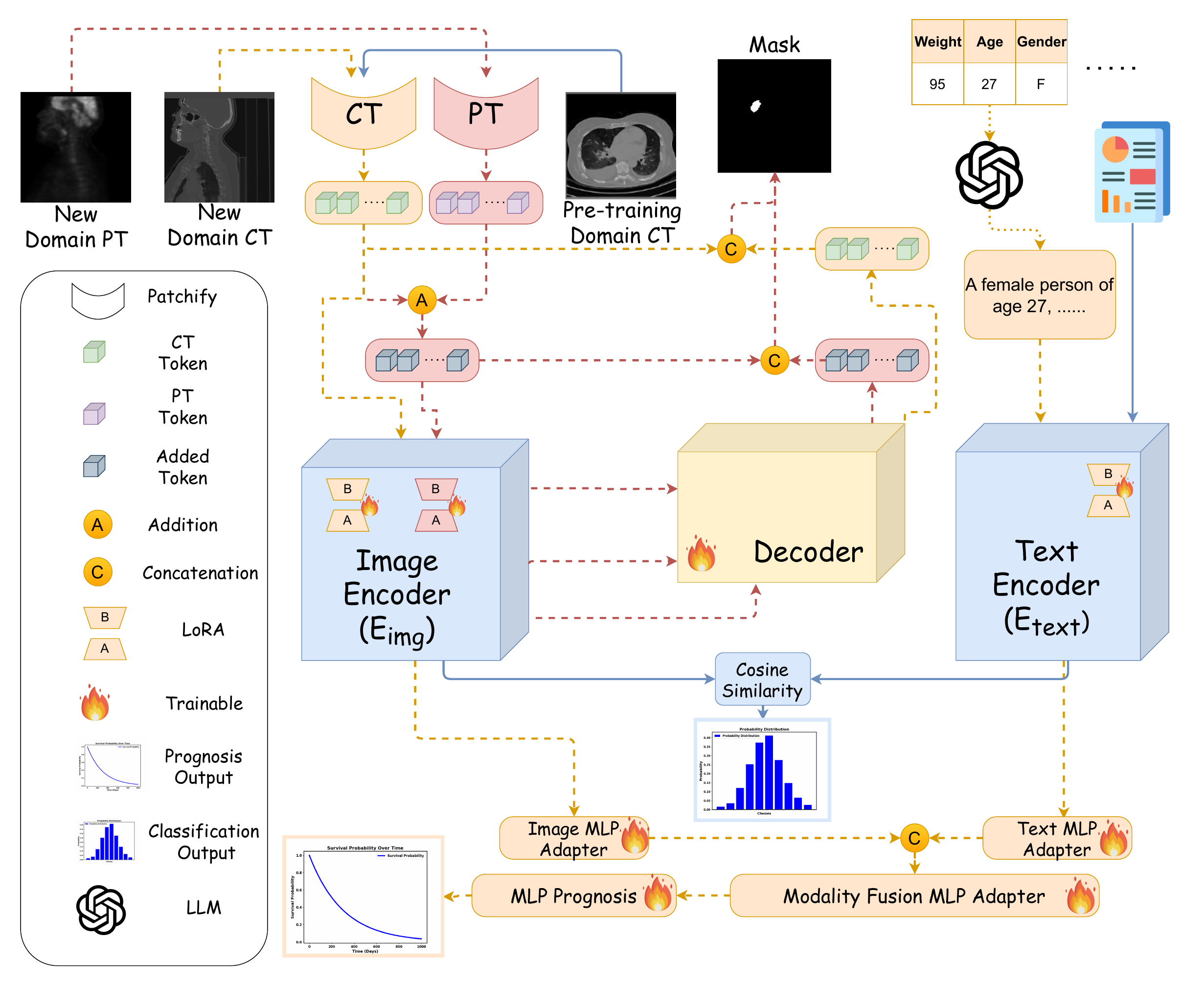}
    \caption{\textbf{Overview of our proposed UNICON framework.} The foundational model (blue blocks) processes pre-training domain CT scans with their corresponding reports to perform classification by computing the similarity between image and text embeddings. The framework introduces three key adaptations: (a) Prognosis Adaptation (Yellow Flow): A large language model (LLM) first restructures EHR data into a report-like format. Then, MLP adapters (image and text), a modality fusion adapter, and LoRA fine-tuning are incorporated to enhance prognosis prediction on the new domain CT scans. (b) Segmentation Adaptation (CT Only): LoRA weights and a decoder are added to enable segmentation using the new domain CT scans alone. (c) Segmentation Adaptation (CT + PET) (Red Flow): The model is further adapted for multimodal segmentation by integrating PT scans of the new domain. This involves adding PT-specific tokens and fusion mechanisms to process both modalities.}
    \label{fig:overall_method}
\end{figure}

\textbf{Problem Statement:} We address the problem of continual learning for medical imaging foundation models, aiming to extend their capabilities across multiple modalities (CT-PET ($CP$) and textual reports ($R$)), tasks (prognosis prediction, segmentation ($M$), etc.), and anatomical regions (abdomen, head-and-neck, etc.) while maintaining the integrity of its foundational knowledge. Given a foundation model with frozen image and text encoders ${\mathbf{E}_{\text{img}}, \mathbf{E}_{\text{text}}}$, the challenge lies in equipping this robust model with the ability to seamlessly adapt to new datasets $\mathcal{X}_{m}$ from diverse modalities and tasks, where $m = 1, 2, \dots, M$, without altering its base parameters. The objective is to enable the adapted model $\mathcal{F}_{\Psi}$ to flexibly map multimodal inputs $(CP, R)$ to corresponding outputs $\mathcal{Y}_{\text{Prognosis}}$ for prognosis and $\mathcal{Y}_{\mathcal{M}}$ for segmentation, transforming a \textbf{static foundation into a dynamic, ever-evolving assistant} for complex medical imaging tasks. All notation terms are defined as follows: $CP$ refers to co-registered CT-PET images, $R$ to corresponding textual reports, and $M$ to segmentation masks. Outputs are denoted as $\mathcal{Y}_{\text{Prognosis}}$ for prognosis labels and $\mathcal{Y}_{\mathcal{M}}$ for segmentation masks. In the unified framework, $\mathcal{Y}_{m}$ represents the output space for any given task $m$.

\subsection{Standard Learning Tasks}


\noindent\textbf{Prognosis Prediction:} Prognosis prediction from multimodal data involves estimating patient outcomes from $CP$ scans and associated $R$ \cite{saeed2021ensemble}. Let $\mathcal{X}_{\text{CP, R}} = \{(x_i, r_i, y_i)\}$, where $x_i$ is a $CP$ volumes, $r_i$ a report, and $y_i$ the prognosis label which could censored i.e., NULL. We define: $\mathcal{F}^{\text{Prog}}_{\Theta_1}: \mathcal{X}_{\text{CP, R}} \rightarrow \mathcal{Y}_{\text{Prog}}$, where $\Theta_1$ denotes the trainable parameters of the prognosis adapter, with embedded tokens $T^{\text{Img}}$ and $T^{\text{Text}}$ integrated through cross-modal attention.\\

\noindent\textbf{Segmentation of CT and PET Volumes:}
Segmentation tasks produce voxel-wise labels $\mathcal{M}_i$ (segmentation masks) for anatomical structures \cite{foster2014review}. For $\mathcal{X}_{\text{CP}} = \{(x_i, \mathcal{M}_i)\}$: $\mathcal{F}^{\text{Seg}}_{\Theta_2}: \mathcal{X}_{\text{CP}} \rightarrow \mathcal{Y}_{\mathcal{M}}$, where $\Theta_2$ denotes the trainable parameters of the segmentation adapter, with patch embeddings $T^{\text{Seg}}$ processed through transformer layers. \\

\noindent\textbf{Classification:} The foundation model in our experiments is initially trained for disease classification from $C$ images and $R$. Let $\mathcal{X}_{\text{C, R}} = \{(x_i, r_i, y_i)\}$: $\mathcal{F}^{\text{Class}}_{\Theta_0}: \mathcal{X}_{\text{CP, R}} \rightarrow \mathcal{Y}_{\text{Class}}$, where $\Theta_0$ represents the base model parameters trained for classification. The challenge lies in extending this base model to new tasks such as prognosis prediction ($\Theta_1$), segmentation ($\Theta_2$), and up to $N$ distinct tasks ($\Theta_1, \Theta_2, \dots, \Theta_N$) across diverse domains without catastrophic forgetting.

\subsection{Unified Continual Learning: UNICON}

Medical imaging foundation models require adaptation to numerous tasks, modalities, and anatomical regions, each necessitating distinct parameter sets $\Theta_1, $ $\Theta_2, $ $\dots, $ $\Theta_N$. To address this, we adopt a parameter-efficient approach where the foundational model $\mathcal{F}_{\Psi}$ is incrementally adapted through lightweight adapters $\mathcal{A}^{(m)}$. The adapters are implemented as either low-rank LoRA modules, non-linear MLP layers, or a decoder module for segmentation reconstruction. Each of these modules is chosen based on the upcoming task.
Our adaptation framework is structured into two main components: within-model adaptation and post-model adaptation.

\noindent \textbf{Within-model Adaptation (WMA).} For this, we leverages LoRA \cite{hu2021lora} to efficiently update pre-trained model parameters. This technique introduces low-rank matrices $\Phi_1 \in \mathbb{R}^{d \times r}$ and $\Phi_2 \in \mathbb{R}^{r \times h}$ into the original weight matrix, resulting in an adapted weight matrix given by: $W' = W + \Phi_1 \Phi_2$, where $r \ll \min(d,h)$ with $d$ representing the input dimension and $h$ the output dimension. This formulation ensures efficient fine-tuning while retaining the core model's learned knowledge.

\noindent \textbf{Post-model Adaptation (PMA).}  In this stage, we focus on refining the features extracted by the pre-trained encoder by employing non-linear MLP layers to learn robust representations from the encoder embeddings. A key component here is the multi-modal fusion MLP layer, which integrates and learns combined representations from multiple modalities. For task-specific requirements, such as segmentation, we also include a dedicated decoder module that reconstructs the segmentation mask from the learned embeddings.

\subsubsection{Resolution Adaptation.}
Pre-trained models are often optimized for a fixed image resolution, which can be problematic in real-world scenarios where scan resolutions vary. To efficiently adapt our model to new modalities or domains, we address this resolution mismatch by dynamically incorporating a patch embedding layer within the transformer architecture for each task. During training, both the patch embedding layer and the positional embeddings, along with the adaptation modules, are fine-tuned. This approach enables the model to seamlessly adjust to different resolutions while maintaining robust performance across various domains.




Figure \ref{fig:overall_method} shows our proposed approach. UNICON unifies continual learning across tasks and modalities by incrementally integrating new adaptation modules at each learning step. Let $\mathcal{X}$ denote the input space encompassing $CP$ and $R$, and $\mathcal{Y}_{m}$ the output space for any task $m$. At each step $m$, $m = 1, 2, \dots, M$, a new adaptation module $\mathcal{A}^{(m)}$ is trained without modifying the frozen base parameters. The unified continual learning function is:
\begin{equation}
\mathcal{F}^{\text{UNICON}}_{\Psi}: \bigcup_{m=1}^{M} \mathcal{X}_{m} \rightarrow \{\mathcal{Y}_{m}\},
\end{equation}
where $\mathcal{X}_{m}$ represents the input data for task $m$. The parameter set $\Psi$ comprises the frozen FM encoders $\mathbf{E}_{\text{img}}, \mathbf{E}_{\text{text}}$, and all adaptation modules $\{\mathcal{A}^{(m)}\}$. Each adaptation module is dynamically selected during inference. Thus:
\[
\Psi = \{\mathbf{E}_{\text{img}}, \mathbf{E}_{\text{text}}\} \cup \{\mathcal{A}^{(m)} : m = 1, 2, \dots, M\}.
\]

\noindent The image encoder $\mathbf{E}_{\text{img}}$ processes $CP$ into latent representations, while $\mathbf{E}_{\text{text}}$ embeds $R$. The adaptation modules $\mathcal{A}^{(m)}$ refine these embeddings for specific tasks and modalities, ensuring knowledge retention and seamless continual learning.

\noindent\textbf{Inference.} Our adaptation modules serve as dynamic extensions to the foundational model, selectively activating the relevant components during inference based on the input and target task. This design enables a model initially trained for specific tasks and modalities to seamlessly extend its capabilities across multiple tasks and modalities. By efficiently routing the data through modality-specific and fusion adapters - as well as the decoder for segmentation - the framework ensures flexible and scalable inference without requiring extensive retraining.

\section{Experimental Setup}

\begin{table}[!t]
\centering
\caption{Progressive adaptation learning sequence in our experiments.}
\label{tbl:exp_main}
\setlength{\tabcolsep}{4pt}
\scalebox{0.8}{
\begin{tabular}{l | c c c c}
\toprule
\rowcolor{Gray}
{\textbf{Order}$\rightarrow$} & {\textbf{Foundational Model}} & {\textbf{Step 1}} & {\textbf{Step 2}} & {\textbf{Step 3}} \\
\midrule
\multirow{1}{*}{\textbf{Body Region}} 
 & Chest $\rightarrow$ & {Head \& Neck} $\rightarrow$ & Head \& Neck & $\rightarrow$ Head \& Neck \\
\multirow{1}{*}{\textbf{Modality}} 
 & CT $\rightarrow$ & CT $\rightarrow$ & CT $\rightarrow$ & CT \& PET \\
\multirow{1}{*}{\textbf{Task}} 
 & Classification $\rightarrow$ & Prognosis $\rightarrow$ & Segmentation $\rightarrow$ & Segmentation \\
\bottomrule
\end{tabular}}
\end{table}

\noindent \textbf{Dataset Description.} We used the publicly available HECTOR dataset (HEad and neCK TumOR) because it is one of the few datasets that comprises multiple modalities i.e., CT and PET scans, segmentation masks, and electronic health records (EHR) from 488 patients collected from seven centers and different machine types \cite{andrearczyk2021overview}. This comprehensive dataset includes Recurrence-Free Survival (RFS) information, detailing both time-to-event outcomes and censoring status. 


\noindent \textbf{Configurations.} In this study, we employ the CT CLIP model as our foundational model \cite{hamamci2024developing}. CT CLIP is pre-trained on CT scans of the chest region for the classification task. Table \ref{tbl:exp_main} details the sequential adaptation of the CT CLIP model across various tasks and domains. Our process begins with adapting the foundation model for the prognosis task, followed by segmentation using only CT scans, and finally extending to segmentation with both PET and CT scans. The task-specific configurations are outlined below.

For patient outcome prediction, we integrate two survival models—DeepHit \cite{lee2018deephit} and MTLR \cite{fotso2018deep}. To tailor the CT CLIP model for survival analysis, we incorporate within-model adapters modules and post-model adapters modules. This approach minimizes parameter overhead while enhancing task-specific fine-tuning. Prognostic performance is assessed using the concordance index (C-index). Models are optimized using the AdamW optimizer with a learning rate of 3e-4 and a weight decay of 1e-5, trained for 50 epochs with a batch size of 16. The CT scans are resized to the resolution accepted by the CT CLIP model, keeping the scale intact. Then, the values are clipped to the range ($-$1024, 1024) and subsequently normalized to [$-$1, 1]. For the text report, the EHR data is converted into a medical report using prompt engineering on the GPT-4 model \cite{achiam2023gpt}. The best model is selected based on the highest validation C-index.

For segmentation, we used the popular UNETR 3D model as a baseline and adapted it \cite{hatamizadeh2021transformers}. The models are trained for a maximum of 25k steps and use the MONAI library \cite{cardoso2022monai}. All the CT and PET scans were resized to 96X96X96 and clipped and normalized.  PET images are pre-processed via Z-score normalization. All segmentation experiments utilize the AdamW optimizer with a learning rate of 1e-4, a weight decay of 1e-5, and a batch size of 1. The model with the best average Dice score in the validation set is chosen as the optimal model.

\section{Results And Discussion}

\begin{table}[!t]
\begin{minipage}{0.45\textwidth}
\centering
\caption{Prognosis adaptation performance with different input modalities using C-Index. We used DeepHit \cite{lee2018deephit} and MTLR \cite{fotso2018deep} survival model performance under different input modalities.}
\label{tbl:prognosis_main}
\setlength{\tabcolsep}{4pt}
\scalebox{0.70}{
\begin{tabular}{l | c c | c c}
\toprule
\rowcolor{Gray}
{} & \multicolumn{2}{c}{\textbf{Adaptation}} & \multicolumn{2}{c}{\textbf{C-Index}} \\
\rowcolor{Gray}
\multirow{-2}{*}{\makecell[l]{\textbf{Input Modality} $\downarrow$}} & \textbf{PMA} & \textbf{WMA} & \textbf{MTLR} & \textbf{DeepHit} \\
\midrule
\multirow{1}{*}{\textbf{Baseline \cite{saeed2024survrnc}}} 
 & \ding{55} & \ding{55} & 0.634 & 0.661 \\
\midrule
\multirow{2}{*}{\textbf{Text}}
 & \ding{51} & \ding{55} & 0.668 & 0.686 \\
 & \ding{51} & \ding{51} & 0.652 & 0.683 \\
 \midrule
\multirow{2}{*}{\textbf{Image}} 
 & \ding{51} & \ding{55} & 0.546 & 0.566 \\
 & \ding{51} & \ding{51} & 0.603 & 0.626 \\
 \midrule
 \multirow{2}{*}{\textbf{Image and Text}}
 & \ding{51} & \ding{55} & 0.658 & 0.698 \\
 & \ding{51} & \ding{51} & 0.670 & 0.721 \\
\bottomrule
\end{tabular}}
\end{minipage}
\hspace{1.5em}
\begin{minipage}{0.45\textwidth}
\centering
\caption{Performance of Segmentation Adaptation with Different Input Medical Modalities in Dice Score (D-Score). We used UnetR \cite{hatamizadeh2021transformers} model for basline and adapted it for the experiments.}
\label{tbl:segmentation_main}
\setlength{\tabcolsep}{4pt}
\scalebox{0.80}{
\begin{tabular}{l | c c | c}
\toprule
\rowcolor{Gray}
{} & \multicolumn{2}{c}{\textbf{Adaptation}} & {} \\
\rowcolor{Gray}
\multirow{-2}{*}{\makecell[l]{\textbf{Medical Modality} $\downarrow$}} & \textbf{PMA} & \textbf{WMA} & \multirow{-2}{*}{\makecell[l]{\textbf{D-Score}}} \\
\midrule
\multirow{1}{*}{\textbf{CT (Baseline) \cite{hatamizadeh2021transformers}}} 
 & \ding{55} & \ding{55} & 0.609\\
\midrule
\multirow{2}{*}{\textbf{CT}} 
 & \ding{51} & \ding{55} & 0.615 \\
 & \ding{51} & \ding{51} & 0.628 \\
\midrule
 \multirow{1}{*}{\textbf{CT \& PET}}
 & \ding{51} & \ding{51} & 0.657 \\
\bottomrule
\end{tabular}}
\end{minipage}
\end{table}

We evaluate the prognostic performance of our adapted CT CLIP model on the HECKTOR dataset using 5-fold cross-validation. Baseline performance is established using two well-known survival models—DeepHit \cite{lee2018deephit} and MTLR \cite{fotso2018deep}. Initially, we adapt the CT CLIP model for the prognosis task by integrating the within-model and post-model adaptation on the text encoder. As shown in Table \ref{tbl:prognosis_main}, these adaptation modules improve the performance of the CT CLIP foundational model compared to the baseline task-specific models, resulting in an approximate 2\% improvement in the C-index. Furthermore, the best results are achieved when adaptation methods are applied simultaneously. In particular, adapting both the text and image encoders using these modules with DeepHit produced the highest C-index of 0.721. It is important to note that adopting only the image encoder did not produce any significant improvement, which can be attributed to the heavy reliance of the prognostic task on the EHR data processed through the text modality.

For segmentation, we trained UNETR \cite{hatamizadeh2021transformers} on the same data and configuration as the baseline. We started by continually adapting the model with only CT scans with a post-model adaptation module, specifically a decoder. Then, we utilized the adaptation module individually and jointly. Table \ref{tbl:segmentation_main} shows the results of the CT adaptation for segmentation. Using just CT gave an improvement of 2\% on the dice score. Furthermore, utilizing PET scans along with CT produced the highest dice score of 65.7\%. Table \ref{tbl:result_tab} presents a comparison between the baseline CT-CLIP model and its UNICON-adapted version. The comparison is based on both the functional capabilities of the models and their performance across different adaptations.

These results highlight several key aspects. First, they demonstrate the effectiveness of leveraging the foundation model as a knowledge base for a wide range of downstream applications. By utilizing the rich representations learned during large-scale pre-training on diverse datasets, our approach efficiently transfers knowledge to new tasks, domains, and modalities—spanning prognosis, classification, and segmentation across CT and PET scans. This knowledge transfer is facilitated by LoRA and adapter modules, which enable efficient adaptation while requiring minimal additional computational resources. This is particularly valuable in medical imaging, where the scarcity of labeled data poses a significant challenge. Second, unlike traditional continual learning, which requires predefined boundaries for adaptation, our approach allows the model to adapt dynamically to new domains and tasks without modifying the foundational model’s parameters. This flexibility is crucial, as real-world sequential data is not strictly confined to changes in class, task, domain, or modality but may encompass any combination of them.

\begin{table}[!t]
\centering
\caption{Comparison of the CT-CLIP and UNICON-adapted version across tasks. This table compares the capabilities of the FM with the UNICON-adapted model across different tasks: Classification (Cls), Prognosis (Prog), Segmentation on CT (Seg (C)), and Segmentation on CT \& PET (Seg (CP)).}
\label{tbl:result_tab}
\setlength{\tabcolsep}{4pt}
\scalebox{0.8}{
\begin{tabular}{l | c c c c}
\toprule
\rowcolor{Gray}
{\textbf{Task}$\rightarrow$} & {\textbf{Cls}} & {\textbf{Prog}} & {\textbf{Seg (C)}} & {\textbf{Seg (CP)}} \\
\midrule
\multirow{1}{*}{\textbf{CT-CLIP}} 
 & \ding{51} & \ding{55} & \ding{55} & \ding{55} \\
\multirow{1}{*}{\textbf{UNICON Adapted}} 
 & \ding{51} & \ding{51} (72.1 C-Index) & \ding{51} (62.8 D-Score) & \ding{51} (65.7 D-Score) \\
\bottomrule
\end{tabular}}
\end{table}

Third, our findings indicate that fine-tuning only a small subset of parameters enables the foundational model to retain its original semantic understanding while effectively adapting to the specific characteristics of new data. This is particularly important in medical imaging, where acquiring large, well-annotated datasets is often infeasible. Finally, this framework effectively addresses domain shift, ensuring that the foundational model remains robust even when applied to new modalities or tasks beyond its original training scope. By continuously fine-tuning and extending the model to incorporate new modalities and tasks, our approach mitigates the risk of catastrophic forgetting and supports incremental learning, making it well-suited for evolving medical applications.



\section{Conclusion}

In this work, we introduced UNICON, a novel continuous training framework that extends the capabilities of foundational medical imaging models beyond the adaptation of a single domain or task. Our approach successfully unifies traditional types of continual learning, enabling a single model to handle multiple modalities and tasks. Our experiments demonstrate that the adapted foundational model performs comparative task-specific models with minimal computational overhead. This work emphasizes the potential of continual learning paradigms in overcoming data scarcity and improving the versatility of medical imaging systems. One potential limitation of this work is to quantify the amount of data required for adapting to a new domain or task. Future research will focus on further refining these techniques and exploring their applicability to a broader range of clinical tasks and imaging modalities.

%
%
%
\bibliographystyle{splncs04}
\bibliography{mybibliography}
\end{document}